\documentstyle[epsf,12pt]{article}
\begin{document}

\catcode`@=11
\long\def\@caption#1[#2]#3{\par\addcontentsline{\csname
  ext@#1\endcsname}{#1}{\protect\numberline{\csname
  the#1\endcsname}{\ignorespaces #2}}\begingroup
    \small
    \@parboxrestore
    \@makecaption{\csname fnum@#1\endcsname}{\ignorespaces #3}\par
  \endgroup}
\catcode`@=12
\newcommand{\newc}{\newcommand}
\newc{\gsim}{\lower.7ex\hbox{$\;\stackrel{\textstyle>}{\sim}\;$}}
\newc{\lsim}{\lower.7ex\hbox{$\;\stackrel{\textstyle<}{\sim}\;$}}
\newc{\gev}{\,{\rm GeV}}
\newc{\mev}{\,{\rm MeV}}
\newc{\ev}{\,{\rm eV}}
\newc{\kev}{\,{\rm keV}}
\newc{\tev}{\,{\rm TeV}}
\newc{\mz}{m_Z}
\newc{\mpl}{M_{Pl}}
\newc{\chifc}{\chi_{{}_{\!F\!C}}}
\newc\order{{\cal O}}
\newc\CO{\order}
\newc\CL{{\cal L}}
\newc\CY{{\cal Y}}
\newc\CH{{\cal H}}
\newc\CM{{\cal M}}
\newc\CF{{\cal F}}
\newc\CD{{\cal D}}
\newc\CN{{\cal N}}
\newc{\eps}{\epsilon}
\newc{\re}{\mbox{Re}\,}
\newc{\im}{\mbox{Im}\,}
\newc{\invpb}{\,\mbox{pb}^{-1}}
\newc{\invfb}{\,\mbox{fb}^{-1}}
\newc{\yddiag}{{\bf D}}
\newc{\yddiagd}{{\bf D^\dagger}}
\newc{\yudiag}{{\bf U}}
\newc{\yudiagd}{{\bf U^\dagger}}
\newc{\yd}{{\bf Y_D}}
\newc{\ydd}{{\bf Y_D^\dagger}}
\newc{\yu}{{\bf Y_U}}
\newc{\yud}{{\bf Y_U^\dagger}}
\newc{\ckm}{{\bf V}}
\newc{\ckmd}{{\bf V^\dagger}}
\newc{\ckmz}{{\bf V^0}}
\newc{\ckmzd}{{\bf V^{0\dagger}}}
\newc{\X}{{\bf X}}
\newc{\bbbar}{B^0-\bar B^0}
\def\bra#1{\left\langle #1 \right|}
\def\ket#1{\left| #1 \right\rangle}
\newc{\sgn}{\mbox{sgn}\,}
\newc{\m}{{\bf m}}
\newc{\msusy}{M_{\rm SUSY}}
\newc{\munif}{M_{\rm unif}}
%
%
\def\NPB#1#2#3{Nucl. Phys. {\bf B#1} (19#2) #3}
\def\PLB#1#2#3{Phys. Lett. {\bf B#1} (19#2) #3}
\def\PLBold#1#2#3{Phys. Lett. {\bf#1B} (19#2) #3}
\def\PRD#1#2#3{Phys. Rev. {\bf D#1} (19#2) #3}
\def\PRL#1#2#3{Phys. Rev. Lett. {\bf#1} (19#2) #3}
\def\PRT#1#2#3{Phys. Rep. {\bf#1} (19#2) #3}
\def\ARAA#1#2#3{Ann. Rev. Astron. Astrophys. {\bf#1} (19#2) #3}
\def\ARNP#1#2#3{Ann. Rev. Nucl. Part. Sci. {\bf#1} (19#2) #3}
\def\MPL#1#2#3{Mod. Phys. Lett. {\bf #1} (19#2) #3}
\def\ZPC#1#2#3{Zeit. f\"ur Physik {\bf C#1} (19#2) #3}
\def\APJ#1#2#3{Ap. J. {\bf #1} (19#2) #3}
\def\AP#1#2#3{{Ann. Phys. } {\bf #1} (19#2) #3}
\def\RMP#1#2#3{{Rev. Mod. Phys. } {\bf #1} (19#2) #3}
\def\CMP#1#2#3{{Comm. Math. Phys. } {\bf #1} (19#2) #3}
\relax
%
%
%
\def\beq{\begin{equation}}
\def\eeq{\end{equation}}
\def\bea{\begin{eqnarray}}
\def\eea{\end{eqnarray}}
%
%
%
\newc{\ie}{{\it i.e.}}          \newc{\etal}{{\it et al.}}
\newc{\eg}{{\it e.g.}}          \newc{\etc}{{\it etc.}}
\newc{\cf}{{\it c.f.}}
\def\smuon{{\tilde\mu}}
\def\neut{{\tilde N}}
\def\char{{\tilde C}}
\def\bino{{\tilde B}}
\def\wino{{\tilde W}}
\def\higgsino{{\tilde H}}
\def\sneut{{\tilde\nu}}
\def\stau{{\tilde\tau}}
\newc\stp{\tilde t}
%
%
%
%
\def\slash#1{\rlap{$#1$}/} 
\def\Dsl{\,\raise.15ex\hbox{/}\mkern-13.5mu D} 
\def\delsl{\raise.15ex\hbox{/}\kern-.57em\partial}
\def\Ksl{\hbox{/\kern-.6000em\rm K}}
\def\Asl{\hbox{/\kern-.6500em \rm A}}
\def\Qsl{\hbox{/\kern-.6000em\rm Q}}
\def\gradsl{\hbox{/\kern-.6500em$\nabla$}}
%
%
%
\def\bar#1{\overline{#1}}
\def\vev#1{\left\langle #1 \right\rangle}
%

\begin{titlepage}
\begin{flushright}
February 2002\\
\end{flushright}
\vskip 2cm
\begin{center}
{\large\bf Bounds on Charged, Stable Superpartners from 
Cosmic Ray Production}
\vskip 1cm
{\normalsize\bf
Mark Byrne, Christopher Kolda and Peter Regan} \\
\vskip 0.5cm
{\it Department of Physics, University of Notre Dame\\
Notre Dame, IN~~46556, USA\\[0.1truecm]
}

\end{center}
\vskip .5cm

\begin{abstract}
Supersymmetric models often predict a lightest superpartner (LSP) which is
electrically charged and stable on the timescales of collider
experiments. If such a particle were to be observed experimentally,
is it possible to determine whether or not it is stable on
cosmological timescales? Charged, stable particles are 
usually considered to be excluded by 
cosmological arguments coupled with terrestrial searches for
anomalously heavy water molecules.
But when the cosmology is significantly altered, as can happen in models
with large extra dimensions, these arguments are in turn significantly
weakened. In this paper we suggest an alternate way to use searches for
superheavy water to constrain the lifetimes of long-lived, charged
particles, independent of most cosmological assumptions. By
considering SUSY production by cosmic rays in the upper atmosphere, we
are able to use current bounds on superheavy water to constrain the
mass scale of squarks and gluinos to be greater than about $230\gev$,
assuming a stable, charged LSP. This bound can be increased, but only
by significantly increasing the size of the initial water sample tested.
\end{abstract}

\end{titlepage}

\setcounter{footnote}{0}
\setcounter{page}{1}
\setcounter{section}{0}
\setcounter{subsection}{0}
\setcounter{subsubsection}{0}

\section{Introduction}
One of the primary goals of the current generation of colliders 
is the discovery of the mechanism by which
the mass scale of the weak interactions is generated and stabilized.
Whether that mechanism is supersymmetry (SUSY), technicolor, extra
dimensions or something not yet considered, it will generically 
require a number of new particle states. One or more of these
particles are often stable, at least on experimental time scales, 
thanks to discrete symmetries in the model or suppressed couplings.  
The archetypes for
such particles are the ``LSP'' (lightest SUSY particle) in the most
SUSY models, and the NLSP (next-to-lightest SUSY particle) in
gauge-mediated models. 

If a new particle produced at a collider lives long
enough to escape the detector, there is no simple way to measure its
lifetime. Of utmost importance will be determining whether the
lifetime is indeed finite or if the particle is stable on cosmological
timescales. It has become standard lore that any stable
remnant of new weak-scale physics must be, at most, weakly
interacting. 
The possibility that the stable particle is electrically
charged or strongly interacting has long been dismissed based on
a combination of astrophysical and cosmological arguments. 
If such a particle, often called a CHAMP~\cite{champ},
exists in abundance in our
galaxy, then a steady flux of CHAMPs would have been captured by the
earth over the course of our planet's lifetime. Such stable, charged 
particles would
form hydrogen-like atoms which could in turn form superheavy water
molecules. These superheavy atoms and molecules would
collect in our oceans in significant abundances.
Searches for superheavy isotopes of water in ocean water have all yielded
negative results~\cite{searches,smith}. These searches have
effectively ruled out charged dark matter at the weak scale.

This result is obviously dependent on the flux of CHAMPs onto the earth.
There are two theoretical avenues for determining this flux.
The first assumes that these relics 
represent the majority component of the galactic dark matter
halo; then their local abundance can be estimated directly from 
astrophysical data. From this we easily deduce the rate at which the CHAMPs
would be captured and accumulate in our oceans. 

The second approach is to use standard cosmological assumptions and
techniques to calculate the relic abundance of the CHAMPs from first
principles. Then using calculations of dark matter clumping, 
a local density can be estimated. These calculations also 
predict far greater abundances of superheavy water than are observed
experimentally, even when the abundance is too small to account for 
the dark matter.

There has been recent interest in whether charged stable relics are
ruled out in more cosmologies. Motivated in part by theories with large
extra dimensions~\cite{extra},  
studies have been done to calculate relic abundances
for the case of extremely low reheat temperatures (as low as an
MeV). The surprising result of these studies is that
particles with TeV masses can still be created during the reheat
process even if the final temperature is around the MeV scale~\cite{reheat}.
When applied to electrically charged relics, a bound of roughly
1~TeV on the relic's mass can be deduced using experimental search
bounds~\cite{newchamp}.

But can we bound, exclude or search for 
stable, charged relics without recourse to cosmology?
In some models, physics  near the TeV scale is so non-canonical
that it is unclear whether existing cosmological bounds apply at all
(see for example the model of Ref.~\cite{hall}).
In this paper, we will approach the same problem but from an
original point of view. We will consider charged relics
created by cosmic rays in the earth's upper atmosphere. In order to be
specific, we will consider SUSY as our template model; however
extension of these bounds to any other model of weak-scale physics
with a stable charged relic would be trivial. 

We will not place a bound on the mass of the SUSY relic itself.
Instead we
will place a bound on the SUSY scale, or more specifically, the masses
of the squarks and gluinos.
Direct production of the charged relic in cosmic ray collisions
may be very small (for example, the relic could be a stau with only
small electromagnetic cross-sections). However,
production of any SUSY state would, after a cascade of decays, 
result in one or
more of the charged LSPs. Thus the production cross-section will
generally  be dominated by production of squarks and gluinos, not the
LSP itself.

None of these results depend in any way on new models which attempt to
explain cosmic rays at the very highest energies.
These results are generated using the usual spectrum of
incident protons and heavy nuclei. Our only assumption is
that this spectrum has remained essentially unchanged over the last
3~billion years of the earth's history.

\section{Cosmic Rays as Particle Colliders}

While the energy frontier in particle accelerators is slowing moving
higher and higher, collisions with center-of-mass energies in the
TeV range have been common throughout the history of
the earth in the form of cosmic rays scattering off atmospheric
nuclei. While the vast majority of these cosmic rays lose their energy
through one or more hard QCD interactions, a small fraction can 
undergo new-physics interactions. With energies ranging up to
$10^{11}\gev$ (in the earth's frame), the incident cosmic rays are capable of
generating center-of-mass energies above $10^5\gev$ when scattering off a
proton. And with the exception of the very highest energies,
the incident cosmic ray energy spectrum is well-measured. 

While these weak interactions in the upper atmosphere
are useless for studying short-lived SUSY states, such states will be 
produced nonetheless. And if R-parity is conserved, those SUSY states will
eventually decay down to the LSP. Thus LSPs are produced in our
atmosphere in standard particle physics processes, independent of
cosmological or astrophysical assumptions

\subsection{Spectrum and Composition}

The all-particle spectrum for cosmic rays is well-measured
up to energies of about $10^{11}\gev$. At energies up to about
$10^6\gev$ the spectrum follows a simple $E^{-2.7}$ power law. Around
$10^6\gev$, the so-called ``knee'' is hit at which point the spectrum
begins falling more dramatically, as $E^{-3.2}$. Finally, at energies
around $10^{9}\gev$ (the ``ankle'') the spectrum
flattens out slightly to $E^{-3}$~\cite{greider,pdg}. 

The composition of the incident cosmic rays is
not as well known. The primary spectrum is composed of a number of
elements including protons, helium, iron, etc. For energies below the
``knee'' ($10^6 \gev $ per nucleus), protons are the most abundant
constituents, representing roughly 50\% to 80\% of the 
spectrum. At the knee, iron nuclei begin to dominate the spectrum;
however, only qualitative details of the composition can be
inferred using, for example, the depth of shower maximum.  
The data seems to indicate that 
there is a relative rise in heavier elements and then a
gradual decrease again as the ``ankle'' is approached~\cite{flyseye}.  
Above $10^9\gev$
per nucleus, lighter elements appear to be more abundant again, consistent
with fragmentation of the heavier elements by the CMBR.  
Extractions from the data 
of the average particle number as a function of 
primary energy vary from 5 to 15 for different
experiments~\cite{greider}. Given this level of ignorance, we 
assume a fixed fraction of protons
(50\%) at energies below $10^{6} \gev$, exponentially decreasing
to a constant 1\% at energies above $10^8\gev$. 

We will also assume that the
remainder of the primary composition (the part that is not protons)
has on average $A$ nucleons/incident nucleus; we will choose $A=13$ as
a typical experimental value~\cite{greider}. We will model these
nuclei as a collection of $A$ loosely bound nucleons, each carrying
$E/A$ of the incident energy $E$ of the entire nucleus; in the
center-of-mass frame, the total energy of the collision, $\sqrt{s}$,
is therefore reduced by a factor $A^{-1/2}$. To be
conservative, we assume that any one nuclei can only participate in a
single hard interaction and we do not consider the interactions of the
daughter nuclei, if any are formed, at all. We find that nuclei
heavier than hydrogen are responsible for only about 10\% of the SUSY
interactions and therefore our result is almost independent of how we
model their interactions. We believe the above choices to be very
conservative. Looser constraints with more protons in the spectrum
could allow our calculated rates to more than double.

\subsection{Atmospheric Cascades}

Protons~\footnote{This discussion is easily generalized to neutrons or
heavy nuclei.} incident on our atmosphere will usually lose their energy
either by QCD or QED processes. 
Typical QED energy loss rates through brehmstrahlung and ionization are 
about 2~MeV/(g/cm$^2$)~\cite{pdg}
so that charged particles lose only a few $\gev$ traversing 15~km
of atmosphere. This is negligible compared to losses in hard QCD 
interactions.

Energetic primary protons have a nuclear interaction length of $80\,$g/cm$^2$,
corresponding to 12 QCD interactions over the depth of the
atmosphere~\cite{pdg}. For simplicity, we only consider the first such
interaction and neglect secondaries produced in the resulting air
showers. That is, we assume that once an incident proton has had a
single hard QCD interaction, its energy is degraded beyond the point
at which it is kinematically possible to create SUSY particles.
In effect, we are discarding the possibility that a secondary
could participate in a SUSY interaction, a choice which again results in a 
conservative bound on the number of SUSY states produced.
A complete analysis, unneccesary here, involves solving
a set of coupled cascade equations with appropriate boundary
conditions, taking into account energy loss processes and 
energy-dependent cross sections. Our method reduces to analyzing a 
fixed-target $pp$ collision with one of the protons at very high energies. 

\section{The Calculation}

Now we present the details of our calculation. After passing
through a distance $ d x $ of the atmosphere
with local density $\rho_{atm}$, the flux of protons with energies
between $E$ and $E+\delta E$ which have
undergone hard QCD interactions and are therefore ``lost'' for SUSY
interactions is simply given by 
$$ \frac{1}{n_{\rm atm}(x)}\,
\frac {d \Phi_p(E)}{d x} = -\Phi_p(E)\, \sigma_{pp}(E) $$
where $\Phi_p$ is the flux of protons measured in
GeV$^{-1}\,$cm$^{-2}\,$s$^{-1}\,$sr$^{-1}$, $n_{\rm atm}$ is the number
density of protons in the atmosphere at a depth $x$, and
$E$ is the energy of the incident proton in the earth's
frame. At the energies under consideration here, the QCD $pp$
cross-section, $\sigma_{pp}$, is roughly 100~mb, independent of energy.
Over that same slice of atmosphere, the total number of SUSY
interactions (integrated over all energies), corresponding to the flux
of LSPs, is
$$\frac{1}{n_{\rm atm}(x)}\,\frac{d \Phi_{\rm SUSY}}{d x} = 
\int dE\,\Phi_p(E)\,\sigma_{{\rm SUSY}}(E)\, .$$
Here $\sigma_{\rm SUSY}$ represents the inclusive cross-section for $pp\to
X_{\rm SUSY}$, independent of whether $X$ is squarks, sleptons, etc.
($\sigma_{\rm SUSY}$ is a function of $s=2m_pE$.)
This system of equations is simple to solve. 
For a spectrum of only protons, the total number of SUSY interactions
in cm$^{-2}\,$s$^{-1}\,$sr$^{-1}$ is then:
$$ \Phi_{\rm SUSY} = \int dE\,\left(\frac{d\Phi_p}{dE}\right) 
\left(\frac{\sigma_{\rm SUSY}}{\sigma_{pp}}\right)$$
where $\Phi_p$ and $\sigma_{\rm SUSY}$ are implicitly functions of $E$.
Accounting for the observation that only a fraction $f_p$ of the
incident cosmic rays are protons, and using our stated assumption that
the remaining cosmic rays contain on average $A$ nucleons, then the
above discussion generalizes to:
$$ \Phi_{\rm SUSY} = \int dE\,\left(\frac{d\Phi_i}{dE}\right) 
\frac{1}{\sigma_{pp}}\left[f_p\,\sigma_{\rm SUSY}(E) +
(1-f_p)\,\sigma_{\rm SUSY}\left({\textstyle\frac{E}{A}}\right)\right]$$
where $d\Phi_i/dE$ is the all-particle incident flux of cosmic rays.

There is another, simpler way to understand the above
calculation. Since we only allow each incident cosmic ray one hard
interaction (and only 1 LSP per SUSY interaction), 
we can simply calculate the probability that that one
interaction will be either QCD or SUSY. Since $\sigma_{\rm SUSY}\ll 
\sigma_{pp}$, the probability of a SUSY interaction is then roughly
$\sigma_{\rm SUSY}/\sigma_{pp}$. That probability is then integrated
over the entire flux of incident cosmic rays to find the flux of LSPs produced.

The only unknown quantity that remains is $\sigma_{\rm SUSY}$. Since
this represents a total cross-section, summed over all possible SUSY
final states, there is considerable room for
model-dependence. However, we can make several simplifying
observations/assumptions. First, because the cosmic rays and the
atmospheric nuclei are baryonic,
strong interactions should dominate the SUSY production
processes. Second, in most ``realistic'' SUSY models, there is an
approximate degeneracy among the strongly-interacting sparticles, that
is, among the squarks and the gluinos. These two statements allow us
to greatly simplify the calculation by only considering superQCD
interactions and by assuming that all squarks and gluinos have a
common mass $M_{\rm SUSY}$.

\begin{figure}
\centering
\epsfysize=3truein
\hspace*{0in}
\epsffile{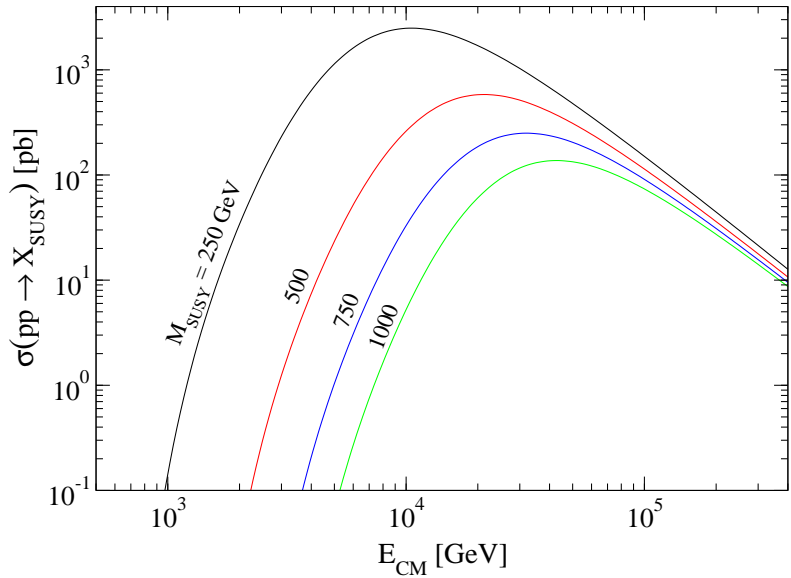}
\caption{Total cross-section for $pp\to\,$SUSY as a function of
$\sqrt{s}$ for several values of $M_{\rm SUSY}$.
}
\label{fig1}
\end{figure}

Independent of the
details, the parton-level cross-section 
for SUSY production obviously goes like $
\alpha_{s}^2/s$. To be slightly more realistic, we will use actual
calculated cross-sections for the range of superQCD processes.
The processes that we consider are:
$$ \begin{array}{lllll}
\mbox{(a)~~}q\bar{q} \rightarrow \tilde{q} \bar{\tilde{q}} & \quad &
\mbox{(c)~~} qq \rightarrow \tilde{q} \tilde{q} & \quad &
\mbox{(e)~~} gg \rightarrow \tilde{g} \tilde{g} \\
\mbox{(b)~~} gg \rightarrow \tilde{q} \bar{\tilde{q}} & &
\mbox{(d)~~} q\bar{q} \rightarrow \tilde{g} \tilde{g} & &
\mbox{(f)~~} qg \rightarrow \tilde{q} \tilde{g} \end{array} 
$$
The relevant cross-sections have been calculated and tabulated 
in Ref.~\cite{beenakker}; we confine our calculation to tree level.
To go to the proton-proton cross-section we
use the CTEQ5M parton distribution functions~\cite{cteq}.
The total $pp\to\,$SUSY cross-section is shown in Figure~\ref{fig1} as
a function of $\sqrt{s}$ for several choices of $M_{\rm SUSY}$.
Over the energy range of interest, the SUSY cross-section is dominated
by $q\bar q\to\tilde{q}\bar{\tilde{q}}, \tilde{q}\tilde{q}$ and
$qg\to \tilde{q}\tilde{g}$. 
There is a subtlety associated with exact versus approximate
degeneracy of the squarks and gluinos and so we consider both the
case in which $m_{\tilde q}-m_{\tilde g}\simeq0$ and $m_{\tilde
q}-m_{\tilde g}\simeq M_{\rm SUSY}$. We find a difference of about 50\%
in our calculated number densities as we vary over this range, and so
we show a range of limits.

Once a squark or gluino has been produced, it will decay in a cascade
down to the the LSP. Presumably, one LSP of positive charge will be
produced for each of negative charge. Searches for the latter
are more difficult and we will concentrate only on the
former. Further, we will assume conservatively that only one
positively charged LSP is produced per SUSY interaction, though the
number can be significantly higher as the cascades of decays progress.
One expects that these positively charged LSPs form superheavy hydrogen
by attracting a nearby electron and that this superheavy hydrogen
eventually (over the lifetime of the earth) bonds into a superheavy
water molecule in the earth's oceans. 

So as a final step we calculate the concentration of
superheavy water in the oceans. Considering the age of the oceans,
$t_O$, to
be roughly 3~billions years, and assuming that the flux of cosmic rays
has remained essentially unchanged over that time period, the number
of superheavy water molecules per usual water molecule in the oceans
(the ``anomalous concentration'', $\chi$) is
$$\chi = 7.6\times 10^{-13}\,
\left(\frac{t_O}{3\,\mbox{Gyr}}\right)
\left(\frac{2.6\,\mbox{km}}{d_O}\right)\times \frac{\Phi_{\rm SUSY}}
{\mbox{cm}^{-2}\mbox{s}^{-1}\mbox{sr}^{-1}}
$$
where $d_O$ is the average depth of the oceans.

\begin{figure}
\centering
\epsfysize=4truein
\hspace*{0in}
\epsffile{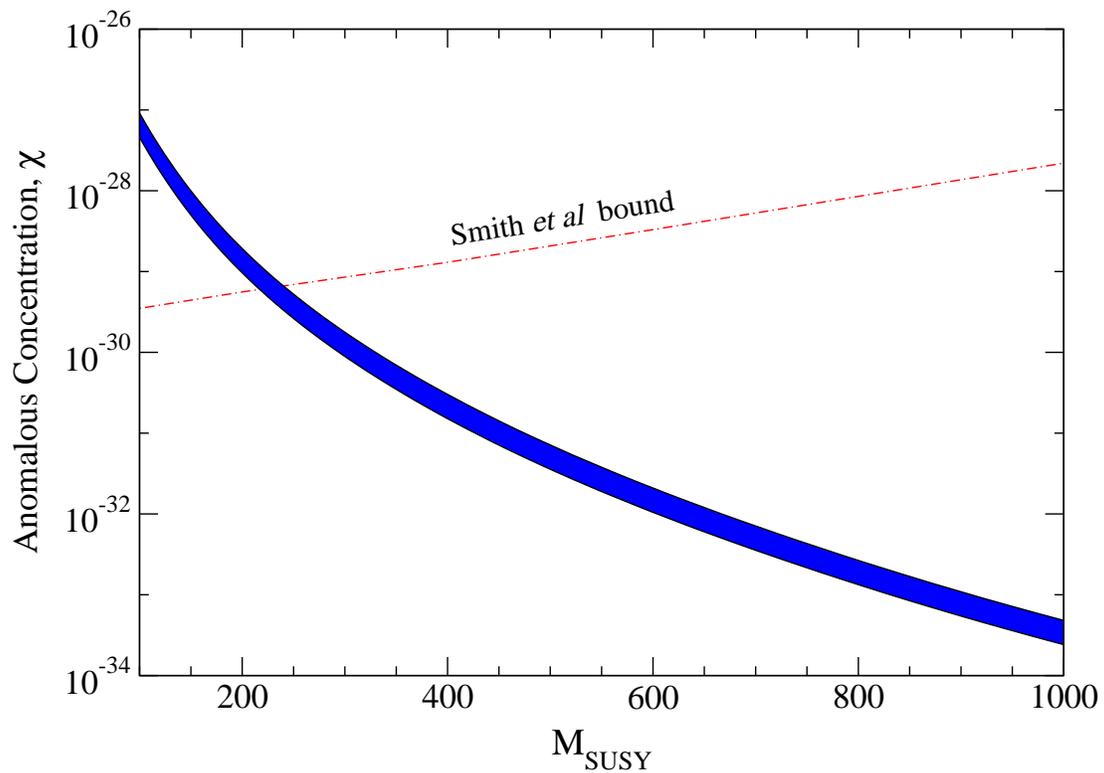}
\caption{Anomalous concentration, $\chi$, of superheavy hydrogen in
sea water as a function of the SUSY mass scale (shaded region). The bound
of Ref.~\cite{smith} is shown as a dashed line. The limits shown here
assume $m_{LSP}=M_{\rm SUSY}$. The width of the shaded region
represents a range of $\pm 50\%$ in the cross-section calculation due
to model-dependence.
}
\label{fig2}
\end{figure}

The constraints on $\chi$ come from searches for superheavy molecules
in large samples of water. Experiments then place bounds on
$\chi$ as a function of the mass of the stable, charged particle. In
our approach, this mass in unknown, though it is bounded from above by
$M_{\rm SUSY}$. The strongest bound
comes the experiment of Smith \etal~\cite{smith},
who find limits ranging from $\chi<3.5\times 10^{-30}$ for $m_{LSP}=100\gev$
to $\chi<8\times 10^{-28}$ for $m_{LSP}=1.2\tev$. Because their bound 
monotonically weakens as $m_{LSP}$ increases, we can place a very
conservative bound on $M_{\rm SUSY}$ by setting $m_{LSP}=M_{\rm
SUSY}$.
For that case, we find $M_{\rm SUSY}> 230\gev$ if the LSP is stable
and charged. (Our bound has an uncertainty of roughly $\pm
10\gev$ due to a $\sim 50\%$ uncertainty that comes from the
subtleties for defining $M_{\rm SUSY}$ discussed above.) 
These results are summarized in Figure~\ref{fig2} where we have shown
the predicted anomalous concentration as a function of $M_{\rm
SUSY}$ and the experimental bound of Ref.~\cite{smith}.

\section{Conclusions}

We have shown that there exist bounds from cosmic ray production in
the upper atmosphere on charged stable relics 
(a charged SUSY LSP in particular) which
are independent of cosmological constraints. Under the assumption that
the incident cosmic ray flux has remained constant over the last 3
billion years, we have calculated a conservative lower bound on the
scale of new physics ($M_{\rm SUSY}$), using nothing more than
standard particle physics. If SUSY has a stable and charged LSP, then
we can place a lower bound on the mass scale of the squarks and gluinos
at 230~GeV.

This procedure can easily be extended to other models of weak-scale
physics. In such a case, limits similar to those found here could be
placed on the masses of new strongly-interacting particles.

Importantly, these bounds will not change if the basic paradigms of
cosmology at and below the weak scale are questioned, such as 
happens in models with large extra dimensions. For example, it is unclear how
standard cosmological bounds can be used to constrain models such as
that of Ref.~\cite{hall} which predicts a light, stable top squark but
becomes non-perturbative and higher-dimensional at the TeV scale. The
bound presented here should hold even in these highly non-standard cases.

Finally, if a charged particle is discovered which is stable on
collider timescales, ruling out or verifying that it is stable on
cosmological timescales will require that searches for superheavy
water be examined again, though with larger initial
samples. Unfortunately, the steeply falling cosmic ray spectrum
requires us to go to exponentially larger samples in order to
significantly increase our sensitivity. For example, the method
used in Ref.~\cite{smith} would require an initial
heavy water sample equal to that contained in the SNO experiment in
order to probe squark masses up to 700~GeV. While such a large-scale
search seems unnecessary at present, any future discovery 
of a new stable, charged
particle might require just such an effort. Otherwise there may be no
other way to study the stability of that state on long timescales.

\section*{Acknowledgements}
CK would like to thank N.~Arkani-Hamed, G.~Domokos, L.~Hall, H.~Murayama and
J.~Poirier for enlightening conversations. This research was supported
by the National Science Foundation under grant PHY00-98791.

\end{document}